\documentclass[preprint,11pt]{elsarticle}

\usepackage{amssymb}

\usepackage{lineno}

\usepackage{color}
\usepackage{listings}
\usepackage{graphicx}
\usepackage{subcaption}
\usepackage{pgfplots}
\usepackage{microtype}

\definecolor{commentgreen}{rgb}{0,0.6,0}

\newcommand{\CC}{C\nolinebreak\hspace{-.05em}\raisebox{.4ex}{\tiny\bf +}\nolinebreak\hspace{-.10em}\raisebox{.4ex}{\tiny\bf +}}
\def\CC{{C\nolinebreak[4]\hspace{-.05em}\raisebox{.4ex}{\tiny\bf ++}}}

\lstdefinestyle{customc}{
  belowcaptionskip=1\baselineskip,
  breaklines=true,
  frame=L,
  xleftmargin=\parindent,
  language=C,
  showstringspaces=false,
  basicstyle=\scriptsize\ttfamily,
  keywordstyle=\color{blue},
  commentstyle=\color{commentgreen},
  stringstyle=\color{orange},
}

\journal{Journal Name}

\makeatletter
\AtBeginDocument{%
	\let\c@figure\c@lstlisting
	
	\let\ftype@lstlisting\ftype@figure 
}
\makeatother

\begin{document}
\begin{frontmatter}



\title{A Traversable Fixed Size Small Object Allocator in \CC}

\author[th,fhg]{Christian Sch\"{u}{\ss}ler}
\author[fhg]{Roland Gruber}

\address[th]{Technische Hochschule N\"{u}rnberg}
\address[fhg]{Department production monitoring, Fraunhofer EZRT}

\address{\{roland.gruber and christian.schuessler\}@iis.fraunhofer.de}


\begin{abstract}
At the allocation and deallocation of small objects with fixed size, the standard allocator of the runtime system has commonly a worse time performance compared to allocators adapted for a special application field.\\
We propose a memory allocator, originally developed for mesh primitives but also usable for any other small equally sized objects.\\
For a large amount of objects it leads to better results than allocating data with the \CC \emph{new} instruction and behaves nowhere worse. The proposed synchronization approach for this allocator behaves lock-free in practical scenarios without using machine instructions, such as compare-and-swap.\\
A traversal structure is integrated requiring less memory than using containers such as STL-vectors or lists, but with comparable time performance.
\end{abstract}


\end{frontmatter}

\section{Introduction}\label{sec:Introduction}
At the allocation and deallocation of huge amounts of small objects, a specialized memory allocator leads to better performance than using the standard allocator of the runtime system.
The purpose of this algorithm is to allocate and deallocate mesh primitives. This means vertices, edges, and faces. A mesh often consists of millions of them. However, this allocator can be applied to any application having the same requirements.\\
We give a brief introduction to the data structures of the implemented memory allocator in the first section.
Afterwards we describe how to preserve the consistency of the traversing data structure in the allocator after a deallocation.
For multithreaded applications a synchronization approach is proposed. The result section shows the time performance and memory consumption of the implemented allocator.

\section{Concepts and Data Structures}\label{sec:ConceptsAndDataStructures}
A common approach for custom memory allocation is to pre-allocate a large memory block and split it into small pieces, which are then assigned to the user. This is often called region-based-memory management  \cite{gay1998memory, grossman2002region, tofte1997region, tofte2004retrospective}. We call a memory block a \emph{bin}, the small pieces corresponding to the individual requested objects are called \emph{chunks}.\\
Allocators that use this approach are considered to be faster than using the default allocator, but it also leads to higher memory consumption \cite{zorn2002reconsidering}. This allocator is designed for a typical chunk size of 32 bytes (for example, four double values). Due to the result section, we suggest a bin size of about 32,000 elements. Therefore a bin consumes about 1 MB including meta data. This amount of memory is negligible in our application field.

Listing~\ref{lst:AllocationDataStructure} shows the internal data structure of a single chunk. The chunk struct consists of a content element (\emph{ContentElement} in line 13) and a status field (\emph{status} in line 22). The most significant bit of the status field marks whether the assigned chunk is free or assigned. The remaining bits are used to save the position of the next free chunk, this data structure is known as \emph{free-list} \cite[p. 36f]{lit:DynamicStorageAllocationASurveyReview}. It is used to get a free chunk for allocation without additional searching for a free element in the memory allocator.\\
If the chunk is free, its memory is interpreted as doubly linked list\,(line 16), storing the next and previous assigned chunk. Otherwise the chunk stores the actual data assigned to the user\,(line 15). The idea to use free chunks to store a doubly linked list is similar to Doug Lea's memory allocator \cite{lit:lea1996memory} and enables a fast traversal over all assigned chunks.\\
The smallest size of a chunk is 20 bytes on a 64-bit system. Two list pointers (16 bytes) and the status field (4 bytes).
An example with a content size of 24 bytes is shown in Figure~\ref{fig:ChunkDataStructure}.
\begin{center}
\vspace{3ex}
\begin{minipage}[t]{0.6\textwidth}
	\includegraphics[width=\textwidth]{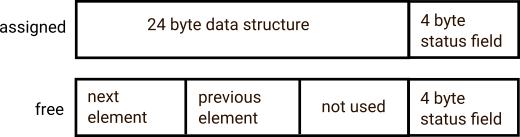}
	\captionof{figure}{The data configuration of an assigned chunk (top). The configuration for a free chunk, with two required list pointers (bottom). In both configurations a 4-byte status field is used.}
	\label{fig:ChunkDataStructure}
\end{minipage}
\vspace{3ex}
\end{center}

\lstset{escapechar=@,style=customc, numbers=left, numberstyle=\scriptsize} 
\begin{lstlisting}[frame=single,caption={internal data structure of \emph{Chunk}},captionpos=b,label={lst:AllocationDataStructure}]
class Chunk;
struct LinkedList
{
  Chunk*  next;
  Chunk*  previous;
};

struct InternalContentType
{
  uint8 content[sizeof(ContentType)];
};

union ContentElement
{
  InternalContentType content;
  LinkedList  linkedList;
};

struct Chunk
{
  ContentElement element;
  uint32 status;
};
\end{lstlisting}
A bin holding several chunks is shown in Figure~\ref{fig:MemoryBin}. 

Only the list entries of free chunks located at the boundary of assigned chunks need to be correct.
These free chunks hold the address of the next assigned chunk and all other free chunks between can be skipped. This approach makes a fast traversal possible because not every single free chunk has to be checked. 
Figure~\ref{fig:MemoryAllocatorFull} shows an example configuration of the memory allocator. All bins are connected through a doubly linked list to enable traversal in both directions.\\
Pseudo elements are used to make a traversal from the beginning to the end of a bin possible. Also they allow us to ignore special cases, such as if there only exists free chunks in a bin.
There is always a chunk marked as assigned at the beginning and at the end of each bin. The chunk right after or before is marked as free. These four pseudo elements are only used by the allocator and are neither visible nor allocatable by the user. Figure~\ref{fig:MemoryAllocatorFull} shows several bins, illustrating this configuration.
\begin{center}
\vspace{3ex}
\begin{minipage}[t]{0.5\textwidth}
	\includegraphics[width=\textwidth]{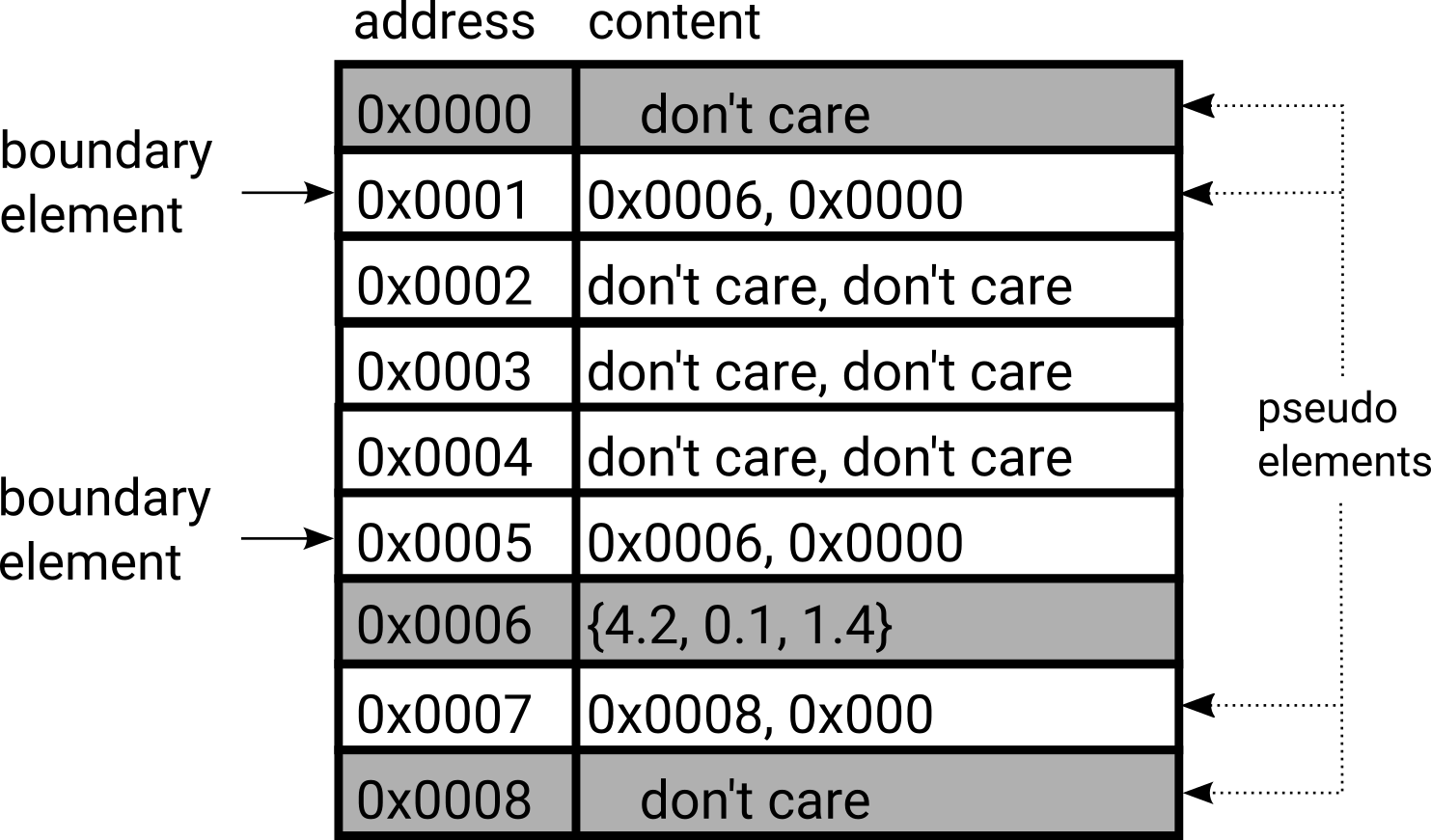}
	\captionof{figure}{In this example bin, an array of float values is stored as user content. The white entries mark free chunks and the gray assigned ones. List entries are only required by free chunks at the boundary to an assigned chunk. The address column is not part of the data structure but is used to illustrate the traversal pointers. Don't care fields mean that the actual content of the chunk is not relevant.}
	\label{fig:MemoryBin}
\end{minipage}
\vspace{3ex}
\end{center}

\section{Traversing Correction Algorithm}
The previous section described the data structure that enables a fast traversal over all assigned chunks. However, after each allocation or deallocation the pointers in the free chunks might be invalid.
For example, if a free chunks gets assigned, the linked list of two free chunks, located before and after, must be corrected to point to this new assigned chunk.\\
In this section we show how the structure can be corrected at an allocation or deallocation call, with time complexity O(1), because we avoid any searching in the data structure.
To make further explanations more clear, some definitions are needed.\\
A pointer in the linked-list of a free chunk is \emph{valid} if it is the \emph{next} member in the linked-list and points to the next assigned chunk or if it is the \emph{previous} member in the linked-list and points to the previous assigned chunk.\\
The structure is \emph{valid} if
\begin{enumerate}
\item The \emph{previous} pointer in the linked-list of every free chunk directly before an assigned chunk is \emph{valid}
\item  The \emph{next} member in the linked-list of every free chunk directly after an assigned chunk is \emph{valid}
\end{enumerate}
If these conditions hold, traversing over all assigned elements is possible.
The structure has to be \emph{valid} again, after each allocation or deallocation call.
To correct the data structure in O(1) time, some further conditions must hold.
\begin{enumerate}
\item In an initialized bin, both pointers in the linked-list in every chunk are \emph{valid}.
\item The free-list is initialized in such a way that elements are assigned in consecutive order.
\item If an element is deallocated, both members in the linked-list in the freed chunk are set valid and the chunk is pushed to the top of the free-list.
\end{enumerate}
Following these conditions, the correction at the allocation and deallocation call can be implemented with only a few if-clauses and write operations. Figure~\ref{fig:MemoryAllocatorFull} shows a valid memory allocator configuration, following all the previous stated conditions.\\
The allocation algorithm is shown in listing \ref{lst:Allocation}.
In the first line, the position of the new element is taken from the free-list.
The position of the chunk, which is taken at the next allocation call, is set in line two. The next two if-clauses are required because there might be free-chunks which are pointing to invalid chunks. This is illustrated in Figure~\ref{fig:AllocationTraversingCorrection} for the correction of the previous pointer.\\
\newpage
\begin{lstlisting}[frame=single,caption={Correction of the structure at allocation},captionpos=b,label={lst:Allocation}]
Chunk* newElement = _baseAddress + _nextFreeElement;
_nextFreeElement = newElement->status & (~FREE_FLAG);
			
if((newElement - 1)->status & FREE_FLAG)
  (newElement->element.linkedList.previous + 1)->element.linkedList.next = newElement;

if((newElement + 1)->status & FREE_FLAG)
  (newElement->element.linkedList.next - 1)->element.linkedList.previous = newElement;
\end{lstlisting}

\begin{center}
\vspace{3ex}
\begin{minipage}[t]{1.1\textwidth}
	\includegraphics[width=\textwidth]{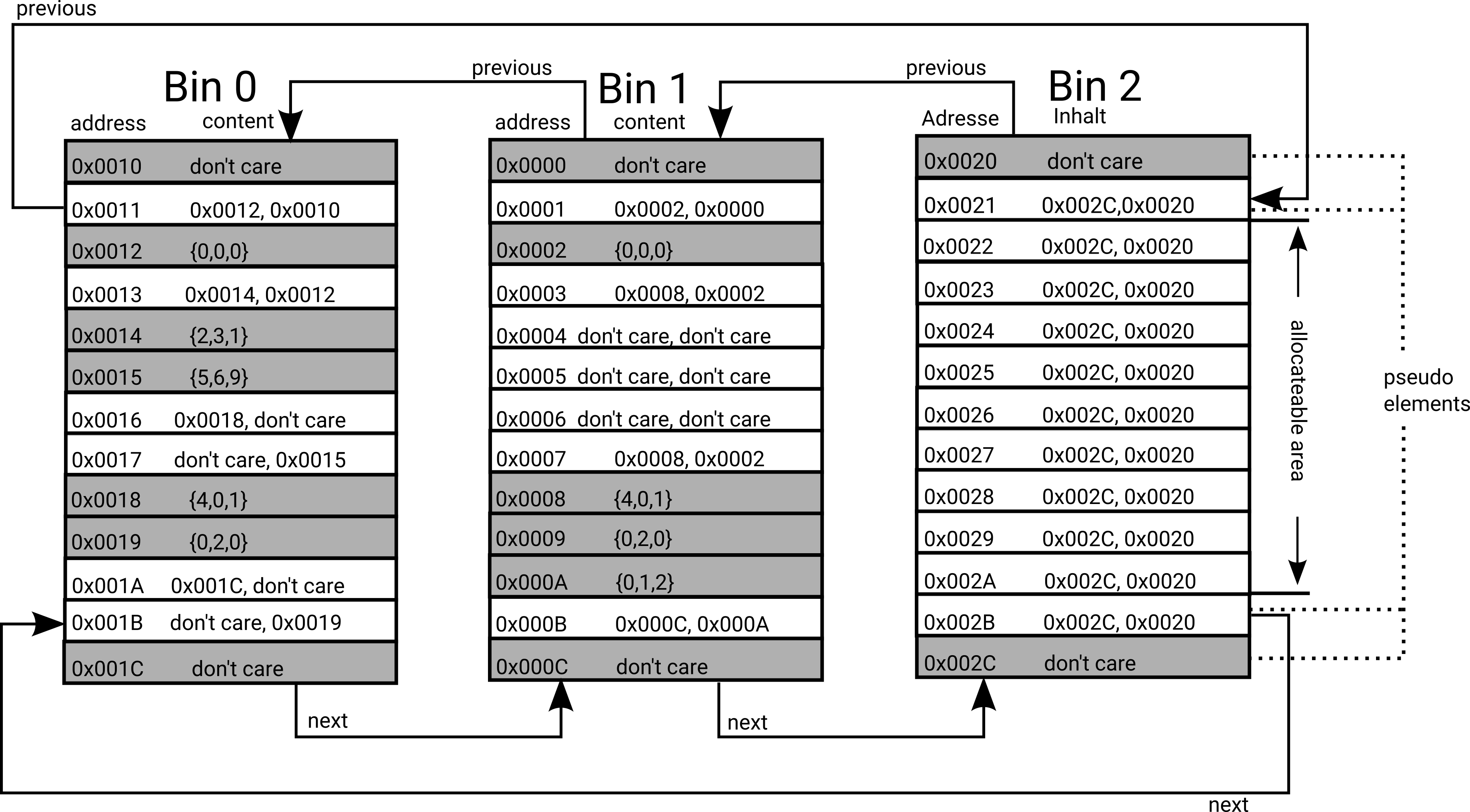}
	\captionof{figure}{Example of the data containers used by the small object allocator. The bins holding the chunks are connected through a doubly linked-list. The right bin represents a new initialized empty bin.}
	\label{fig:MemoryAllocatorFull}
\end{minipage}
\vspace{3ex}
\end{center}

\begin{center}
\vspace{3ex}
\begin{minipage}[t]{0.8\textwidth}
	\includegraphics[width=\textwidth]{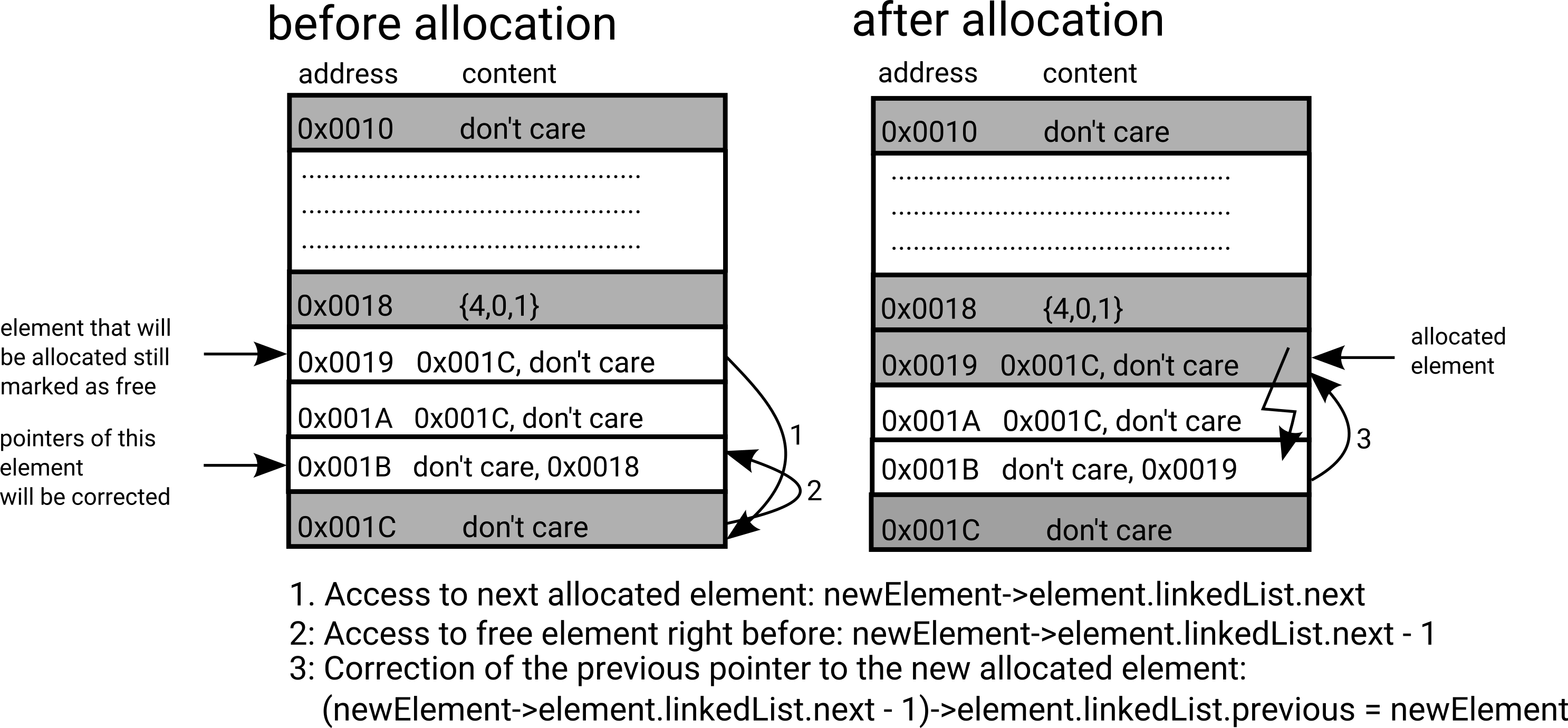}
	\captionof{figure}{Correction of the previous pointer to the new assigned chunk}
	\label{fig:AllocationTraversingCorrection}
\end{minipage}
\vspace{3ex}
\end{center}

For deallocation, the structure has to be corrected with two if-clauses exactly as in the allocation call, to leave the structure valid.
The deallocated chunk has to be marked as free and pushed to the top of the free-list, so it can be assigned at the next allocation call, see line 1-2 in Listing~\ref{lst:Deallocation}. 
The next two conditions in lines 4-12 are necessary to make both pointers in the now free chunk valid. This is because the linked-list of a free chunk assigned in an allocation call has to be valid, see lines 4-8 in Listing~\ref{lst:Allocation}.\\

\begin{lstlisting}[frame=single,caption={Correction of the data structure at deallocation.},captionpos=b,label={lst:Deallocation}]
chunk->status = _firstFreeElement | FREE_FLAG;
_firstFreeElement = static_cast<uint32>(chunk - _baseAddress);

if ((chunk + 1)->status & FREE_FLAG)
  chunk->element.linkedList.nextElement = (chunk + 1)->element.linkedList.nextElement;
else
  chunk->element.linkedList.nextElement = (chunk + 1);

if ((chunk - 1)->status & FREE_FLAG)
  chunk->element.linkedList.previousElement = (chunk - 1)->element.linkedList.previousElement;
else
  chunk->element.linkedList.previousElement = (chunk - 1);
\end{lstlisting}

Correcting the structure in that way, leaves it valid after each allocation or deallocation.
This is shown in the following.\\
After initialization of a bin, it is valid and also every chunk, see Figure~\ref{fig:StateTraversing}. After the first object has been assigned, the previous and the next pointer of the chunk before are both valid, as well as the next pointer of the following chunk. This is sufficient for a valid data structure. A deallocation of this element would lead to the same state in the data structure as when it was initialized, because after a deallocation call, both pointers are made valid.\\ 
At the allocation of all elements after the first one\,(lowest memory address), only the free chunk located after this element has to be valid, because the previous chunk is always assigned if no deallocation appeared.\\ However, the user can deallocate chunks in arbitrary order. In this case the last deallocated chunk will be set valid and its pointer will be pushed on top of the free-list. If this deallocated object is assigned again, the valid linked-list can be used to correct the structure as already shown in listing \ref{lst:Allocation}. The problem is that after the second deallocation, the pointers in the linked list of the deallocated chunk before are getting invalid. So it might be assumed, that this element can never be used to correct the traversing data structure. This problem can be avoided by allocation in reverse order to deallocation. Therefore an allocation leads to exactly that state in which this chunk was valid. So both pointers in its linked list point to the nearest assigned chunk, as explained in the beginning of this section.\\
An example of arbitrary allocation and deallocation calls is shown in Figure~\ref{fig:StateTraversing}.

\newpage
\begin{center}
\begin{minipage}[t]{1.0\textwidth}
	\includegraphics[width=\textwidth]{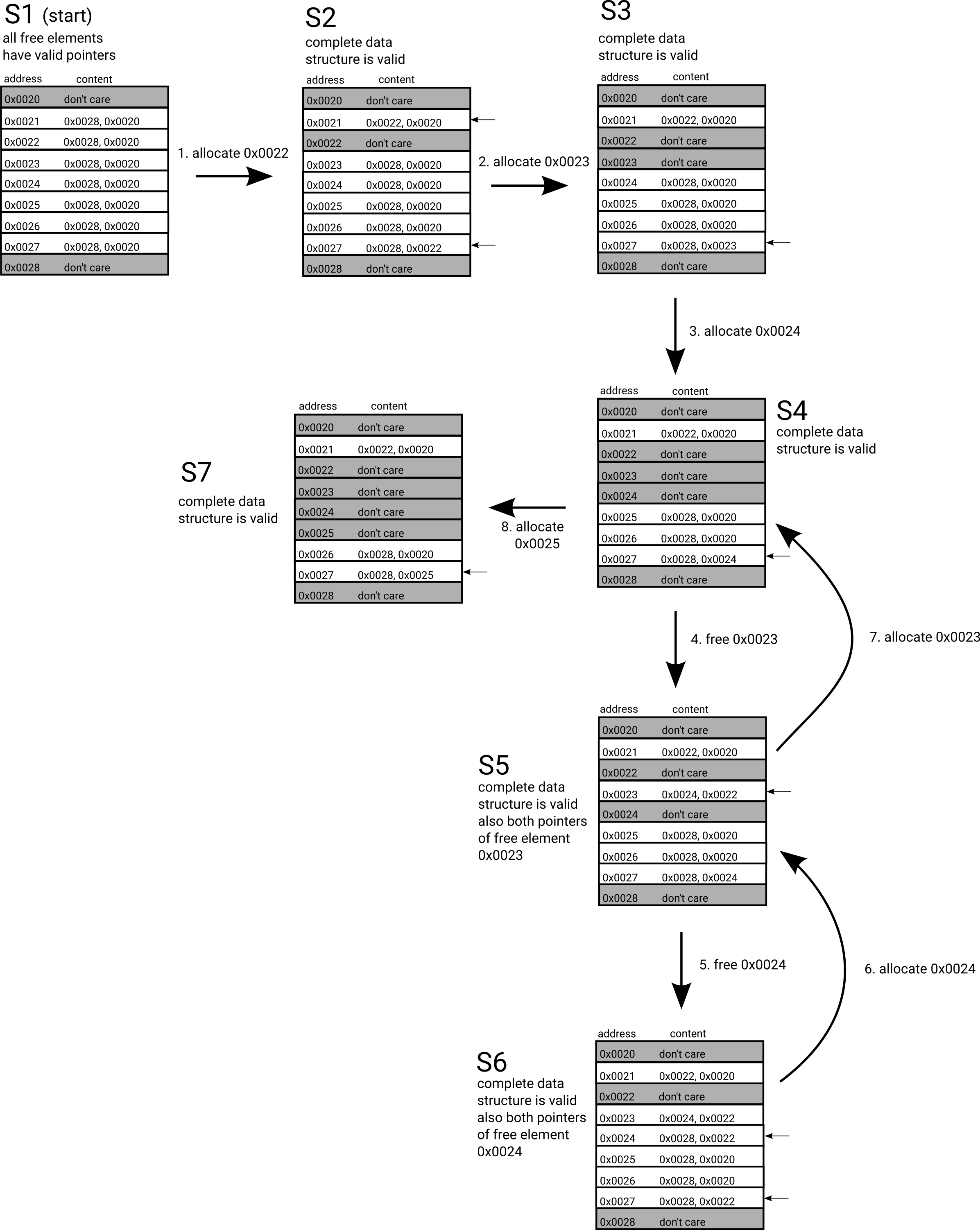}
	\captionof{figure}{Some allocation and deallocation calls changing the state of the data structure. The arrows mark a deallocation or allocation call. The Symbol {\texttt{SX}}, where {\texttt{X}} is a number, marks the current state with description. The small arrows pointing to chunks in a bin mark where the content of a chunk has changed.}
	\label{fig:StateTraversing}
\end{minipage}
\end{center}
\newpage

\section{Thread-Safe-Allocation using readers-writers-locks}\label{sec:Synchronization}
In this section, we propose a synchronization approach that behaves lock-free in most practical applications.\\
The idea is to synchronize the doubly linked-list of bins by using a readers-writers-lock and lock each bin separately using a mutex. The current thread checks for each bin if it is full or locked and gets to the next bin if necessary. Also every thread remembers the last bin where it has assigned a chunk. Therefore the most common case is that every thread allocates memory in a different bin. In this case, allocation works without blocking. Only if a new bin has to be inserted, the complete list must be locked. The probability of this event drops with increasing bin size.
Assigning each thread to a local bin is commonly known as \emph{thread local allocation} \cite{berger2000hoard}. The complete allocation process is summarized in Figure~\ref{fig:AllocationDiagram}.
Our small object allocator has a complexity of $O(n)$ at allocation, due to traversing other bins if necessary. 
This is negligible if the ration of number of allocated chunks to bin size is large enough.\\
Memory can only be completely deallocated if all chunks in a bin are marked as free.\\
In \cite{aigner2015fast} a similar concept for non fixed size elements, using lock-free singly-linked-lists, is proposed. We avoid the usage of completely lock-free data structures, because of the need of machine dependent compare-and-swap instructions and the dealing with the ABA-problem \cite{lit:TheArtOfMultiprocessorProgramming}.

\begin{center}
\begin{minipage}[t]{0.7\textwidth}
	\includegraphics[width=\textwidth]{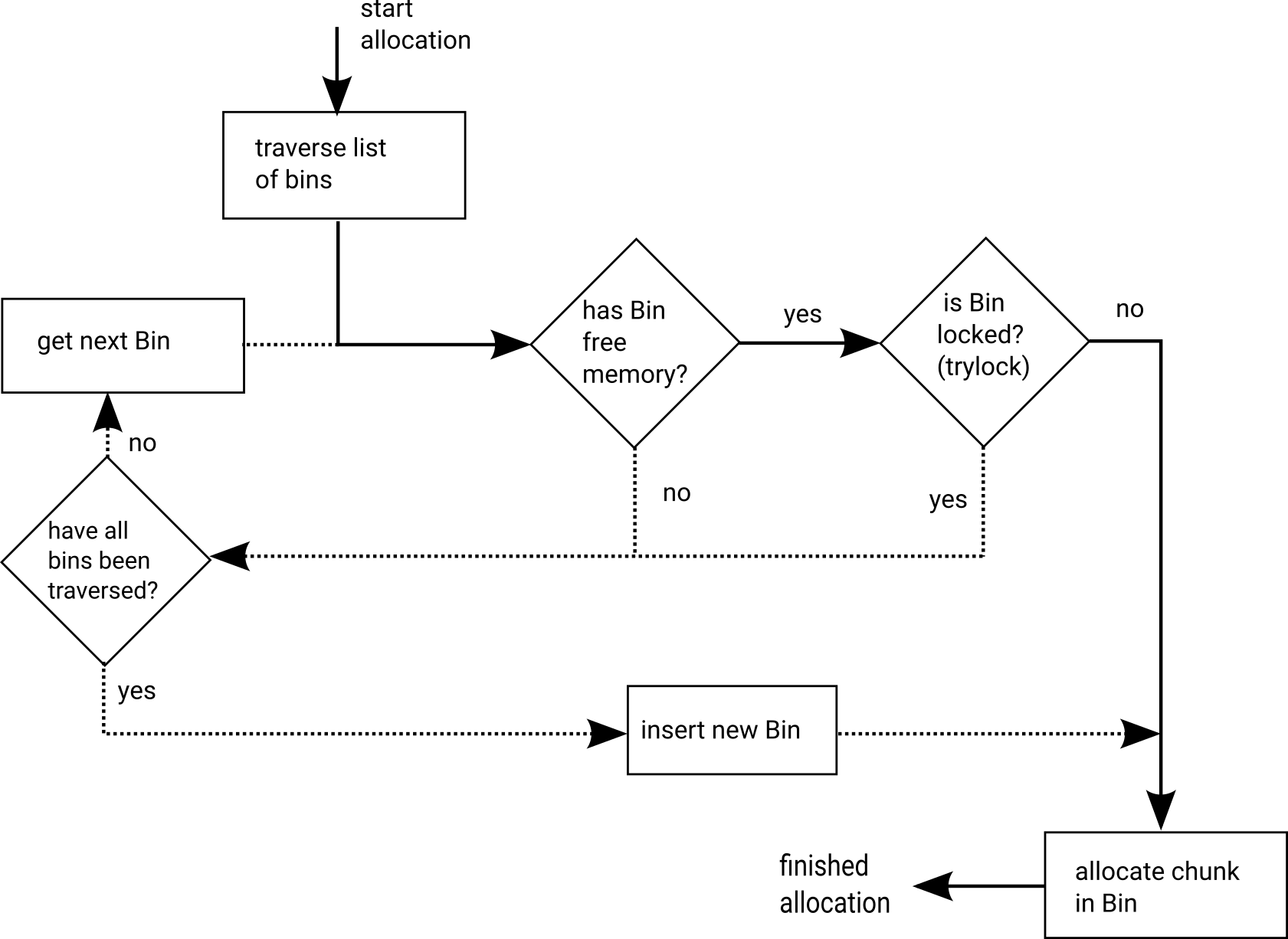}
	\captionof{figure}{Complete allocation scheme for the memory allocator. The solid line marks the most likely path.}
	\label{fig:AllocationDiagram}
\end{minipage}
\end{center}

\section{Multi-Thread Iteration}
The bins in the memory allocator can be distributed into several bin lists, which can be iterated independently.
Each bin list can be assigned to a single thread, so that the iteration speed should be increased by the number of threads. Each begin-iterator of a bin list stores the first assigned chunk in the first bin. Each end-iterator stores the last chunk in the last bin. Figure~\ref{fig:MultipleIterators} gives an example for distributing the bins into two lists.

\begin{center}
	\begin{minipage}[t]{0.7\textwidth}
		\includegraphics[width=\textwidth]{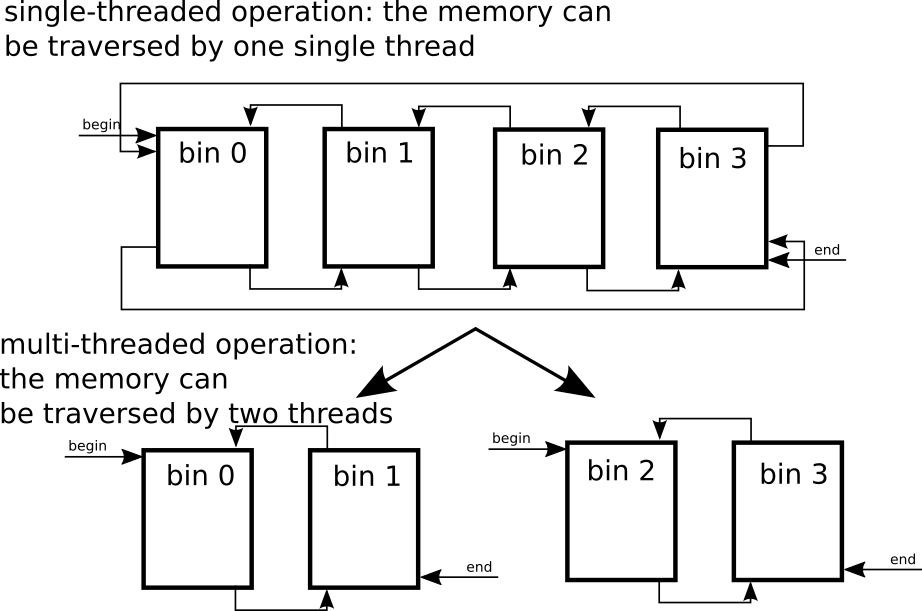}
		\captionof{figure}{The memory allocator is divided into several bin lists. Each pair of begin and end-iterator is assigned to a single bin list. This division allows for all elements in the memory allocator to be iterated independently.}
		\label{fig:MultipleIterators}
	\end{minipage}
\end{center}

\newpage
\section{Results}
Other small object allocation approaches can be found in \cite{lit:bulka2000efficient, coplien1997advanced, lit:modernCppDesign,  lit:stroustrup1986c++, lit:labrosse1998microc}. The boost \emph{memory pool} uses the same concepts as mentioned in \cite{lit:boostOhterImplemenations} and guarantees a working implementation in a well tested library. Therefore it was decided to compare the small object allocator with this implementation.\\
The data type used for the small object is an array of three unsigned integer(64-bit) values. This data type has a typical size of 24 bytes for which the allocator was designed.\\
The test system specifications are stated in Table \ref{table:HardwareSpecifications}.
\begin{table}[h!]
\centering
\begin{tabular}{|l | l |}
\hline
processor & Intel Core i7-4770 CPU @ 3.40 GHz\\
          & 4 cores, hyperthreading switched off\\ 
main memory & 16 GB\\
operation system & Windows 7 Enterprise\\
\hline 
\end{tabular}
\caption{Hardware specifications of the test platform}
\label{table:HardwareSpecifications}
\end{table}

The first benchmark was to allocate 200 million instances of this integer-array type. 
Each allocator was called in a simple for-loop for each object instance. The small object allocator was tested with different numbers of chunks as bin size. The result is shown in Figure~\ref{fig:AllocationDifferentBinSize}. \\
In all following tests the bin size was set to 64,000 chunks.
At this bin size the allocator reached its optimum and is almost as fast as the boost implementation. A fairer way is to compare the allocator with the additional use of STL-containers by the boost memory pool, to make an iteration also for these allocators possible. In this case the small object allocator is clearly faster.\\
The next benchmark measures the performance with concurrent accesses through multiple threads, this is shown in Figure~\ref{fig:AllocationMultiThreaded}. In this case an OpenMP for-loop with different numbers of threads was used. The synchronization technique proposed in section \ref{sec:Synchronization} is much faster than normal mutex locking and behaves like a non-blocking synchronization.\\
Table \ref{table:MemoryConsumption} shows the memory consumption of the complete process. The implemented allocator has the best results. It does not need an additional traversal structure. However, the boost memory pool and the default \CC \  new operator do not differ a lot.\\
The results of the benchmark for the implemented iterators, using the integrated traversal structure are shown in Table \ref{table:IterationTime}. In every case 200 million elements were iterated. Having no fragmentation (no gaps between the assigned chunks) the implemented allocator leads to a performance that is between the STL-vector and the STL-list. 
Additionally, memory gaps were generated in deallocating an element randomly, with the likelihood of 10 and 50\%. Deallocation at random positions is the worst case scenario for the implemented traversal, because unallocated memory has to be skipped and therefore some extra checks have to be done. A more common case is to free a bunch of elements locally near to each other in memory. This behavior should lead to a better time performance.\\
We also benchmarked the parallel iteration with 1 to 8 threads, as shown in Figure~\ref{fig:ParallelIterators}. 
Each iterator executes the code in Listing~\ref{lst:IteratorTask}, which simulates a time-consuming algorithm.
The image shows that the parallel iteration scales well up to 4 threads, which corresponds with the number of cores in the test environment. Cache-misses were examined by measuring all requests that missed the L2 cache \cite{lit:IntelSoftwareManual}, see Figure~\ref{fig:FalseSharingParalelleIteratoren}.
There is no hint of false-sharing, because the number of cache-misses does not differ drastically between the different measurements.

\vspace{3ex}
\makebox[0.9\textwidth][c]{ 
\begin{minipage}{0.52\textwidth}
	\begin{tikzpicture} 
	\begin{axis}[ylabel=time in seconds, xlabel=bin size in elements, legend style={font=\tiny, at={(1.0,1.1)}, anchor=south east}, xtick={4000,8000,16000,32000,64000,128000}, ymax=10.0, ymin=0.0, width=\textwidth, xmode=log, log basis x={2}, font=\tiny, log ticks with fixed point]
	
	\addplot [mark=*,mark options={fill=black,scale=1}] file {results/AllocationVaryBinSizeBoostPool.txt};
	\addplot [mark=*,mark options={fill=white,scale=1}] file {results/AllocationVaryBinSizeBoostPoolList.txt};
	\addplot [mark=x,mark options={fill=black,scale=1}] file {results/AllocationVaryBinSizeBoostPoolVector.txt};
	\addplot [mark=triangle*,mark options={fill=white,scale=1}] file {results/AllocationVaryBinSizeNew.txt};
	\addplot [mark=triangle*,mark options={fill=black,scale=1}] file {results/AllocationVaryBinSizeSmallObjectAllocator.txt};

	\legend{
	boost memory pool,
	boost memory pool with stl list,
	boost memory pool with stl vector,
	c++ new,
	small object allocator,
	}
	\end{axis}
\end{tikzpicture}	

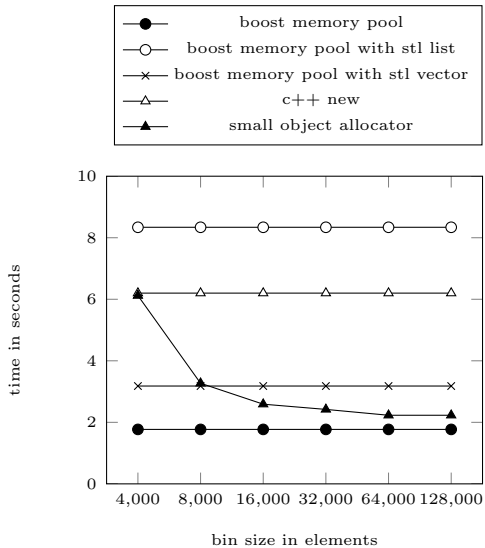
\captionof{figure}{Allocation using the implemented small object allocator with different bin sizes compared to boost memory pool and default new operator of the C++ language. All allocators beside ours were measured once but were plotted for each bin size. Because they have no bin size as parameter.}
\label{fig:AllocationDifferentBinSize} 
\end{minipage}
\hspace{0.1\textwidth}
\begin{minipage}{0.52\textwidth}
	\begin{tikzpicture} 

	\begin{axis}[ylabel=time in seconds, xlabel=number of threads, legend style={font=\tiny, at={(1.0,1.1)}, anchor=south east}, xtick={1,2,4,8,16}, ymax=11.0, width=\textwidth, xmode=log, log basis x={2}, ymin=0.0, font=\tiny, log ticks with fixed point]
	
	\addplot [mark=*,mark options={fill=black,scale=1}] file {results/AllocationBoostPoolMutex.txt};
	\addplot [mark=x,mark options={fill=black,scale=1}]  file {results/AllocationSmallObjectAllocatorMutex.txt};
	\addplot [mark=*,,mark options={fill=white,scale=1}] file {results/AllocationThreadSafeSmallObjectAllocator.txt};

	\legend{
	boost memory pool mutex,
	small object alloctor mutex ,
	small object allocator readers-writers-lock,
	}
	\end{axis}
\end{tikzpicture}	

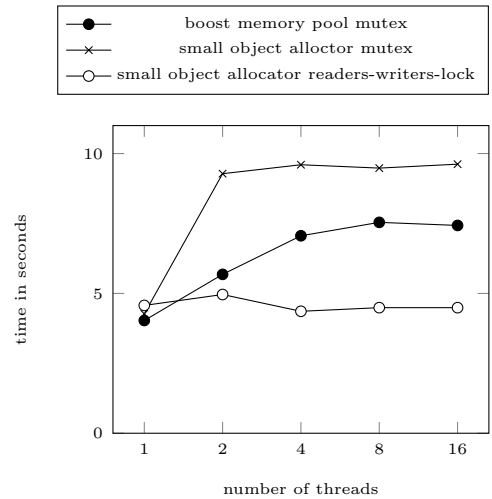
\captionof{figure}{Allocation using the boost memory pool and the small object allocator with different synchronization approaches.}
\label{fig:AllocationMultiThreaded} 
\end{minipage}
}

\vspace{3ex}
\makebox[0.9\textwidth][c]{ 
\begin{minipage}{0.52\textwidth}
	\begin{tikzpicture} 
	\begin{axis}[ylabel=time one thread / time x threads, xlabel=number of threads, legend style={font=\tiny, at={(1.0,1.1)}, anchor=south east}, xtick={1,2,4,8}, ymax=4.0, ymin=0.0, width=\textwidth, xmode=log, log basis x={2}, font=\tiny, log ticks with fixed point]
	
	\addplot [mark=*,mark options={fill=black,scale=1}] file {results/AllocatorParallelIteratros.txt};

	\legend{
	small object allocator iterator speedup,
	}
	\end{axis}
\end{tikzpicture}	

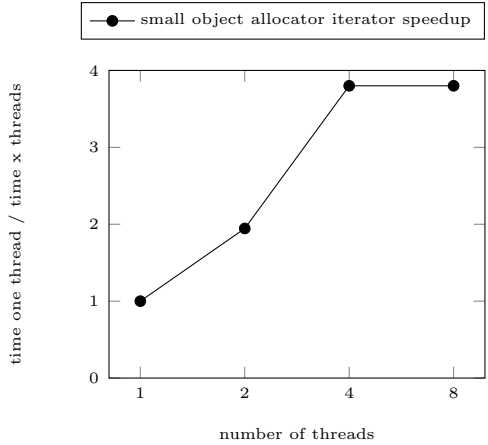
\captionof{figure}{Performance benefit by using parallel iteration at a memory bin size of 64,000 elements.}
\label{fig:ParallelIterators} 
\end{minipage}
\hspace{0.1\textwidth}
\begin{minipage}{0.52\textwidth}
	\begin{tikzpicture} 
	\begin{axis}[ylabel=register value in L2\_RQSTS.MISS in thousand, xlabel=number of threads, xtick={1,2,4,8}, font = \tiny, legend style={font=\tiny, at={(1.0,1.1)}, anchor=south east}, ymax=10000, width=\textwidth, xmode=log, log basis x={2}, log ticks with fixed point];
		
	\addplot file {results/IterationParallelCacheMisses.txt};	
	\legend{
	small object allocator iterator cache misses,
	}
	\end{axis}
\end{tikzpicture}

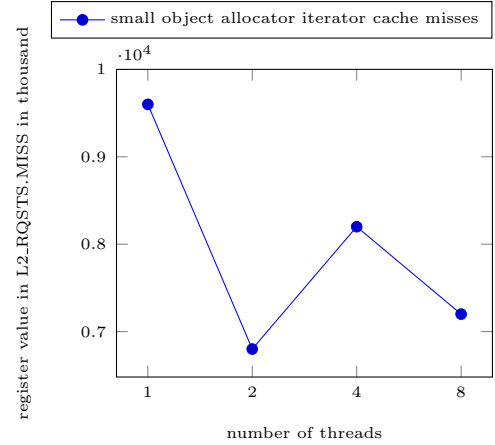
\captionof{figure}{Cache-misses for different numbers of threads.}	\label{fig:FalseSharingParalelleIteratoren} 
\end{minipage}
}
\vspace{0.2\textwidth}

\begin{table}[h!]

	\begin{tabular}{lll }
		Allocator & \hspace{0.05\textwidth} & Memory consumption in MB\\
		\hline
		boost memory pool & \hspace{0.05\textwidth}& \hspace{1mm} 6,308\\
		boost memory pool with vector &\hspace{0.05\textwidth} & \hspace{1mm} 7,876\\
		boost memory pool with list &\hspace{0.05\textwidth}&12,584\\
		\CC \ new & \hspace{0.05\textwidth}&  \hspace{1mm} 6,279\\
		small object allocator \hspace{0.05\textwidth} &\hspace{0.05\textwidth} & \hspace{1mm} 6,276\\
		\hline
	\end{tabular}
	\captionof{table}{Memory consumption comparing different allocator types.}
	\label{table:MemoryConsumption}

\end{table}

\begin{table}[h!]

	\begin{tabular}{lll }
		Container& \hspace{0.1\textwidth} &Iteration time in seconds \hspace{0.05\textwidth} \\
		\hline
		STL-vector & \hspace{0.1\textwidth} & 0.64 \\
		STL-list &\hspace{0.1\textwidth} &  1.15\\
		small object allocator &\hspace{0.1\textwidth}&  0.64\\
		small object allocator (10\% gaps) & \hspace{0.1\textwidth} & 0.79\\
		small object allocator (50\% gaps) & \hspace{0.1\textwidth} &  1.84\\
		\hline
	\end{tabular}
	\captionof{table}{Iteration time using different container classes.}
	\label{table:IterationTime}
\end{table}

\begin{lstlisting}[frame=single,caption={A small benchark applied for every element in the memory allocator.},captionpos=b,label={lst:IteratorTask}]
const std::uint32_t RANDOM_VALUES = 20;

std::uint32_t * randomValues = new std::uint32_t[RANDOM_VALUES];
srand(clock());

for(std::uint32_t i = 0; i < RANDOM_VALUES; i++)
  randomValues[i] = rand() % 10;

for(std::uint32_t i = 0; i < RANDOM_VALUES; i++)
{
  it->x += randomValues[i];
  it->y += randomValues[i] * 2;
  it->z += randomValues[i] * 4;
}
delete[] randomValues;
\end{lstlisting}

\newpage
\section{Summary}
We presented an allocator for fixed sized small objects that allows fast traversal over its elements.\\
The performance of the implemented allocator is comparable with the widely used boost memory pool at single-threaded applications. 
In applications where traversal is required and low memory consumption is favorable, the implemented allocator is superior. 
Additionally, the proposed synchronization approach leads to a non-blocking behavior in practical applications. 
The allocator is very well behaved for parallel iteration.

\section{Acknowledgements}
 I am grateful to J\"org Arndt for several reviews and for supervising my bachelor thesis \cite{lit:SchuesslernMemoryAllocator} that led to this publication. I also want to thank Edith Parzefall for proofreading.




\hspace{0.3\textwidth}
\bibliographystyle{model1-num-names}
\bibliography{bibliography}







\end{document}